# Strongly modulated ultrafast demagnetization and magnetization precession dynamics in ferrimagnetic $Gd_x(CoFe)_{1-x}$ alloys via 3*d*-4*f* intersublattice exchange coupling

Y. Ren[1], L. L. Zhang[1], X. D. He[1], G. J. Wu[2], J. W. Gao[1], P. Ran[1], L. Z. Dan, T.Wang[1], X. W. Zhou[1], Z. Liu[1], J. Y. Xie[3], Q. Y. Jin[2], Zongzhi Zhang[2]

**ABSTRACT:**

Manipulation of the intersublattice interaction strengh ($J_{RE-TM}$) in rare earth (RE)-transition metal (TM) alloys is a key issue to understand how efficiently the laser-induced angular momentum transfers from 3*d* to 4*f* spins and to have a better control of the ultrafast spin dynamics. In this work, the relationships between laser-induced demagnetization process and the intersublattice 3*d*-4*f* interaction for the GdCoFe alloys were systematically studied. The ultrafast two-stage demagnetization process could change into a one-stage mode as the angular momentum transferring channel between 3*d* and 4*f* spins is switched off, which could be modulated by $J_{RE-TM}$. Furthermore, both the effective *g*-factor and damping constant deduced by the subsequently laser-induced magnetization precession process diverge at the angular momentum compensation point based on the ferromagnetic resonance method with the LLG equations. The results provide an alternative way to efficiently manipulate the ultrafast demagnetization time for practical applications.

**Introduction**

Femtosecond (fs) laser-induced spin dynamics of ferromagnetic materials provides an effective way to manipulate spin orientations within the time regime of sub-picosecond (ps), which may have the potential applications in ultrafast magnetic writing and break through the limitation of magnetic recording writing speed (within ~ 100 ps)[1-3]. It was known that the ferromagnetic transition metals (TM) usually present a sub-ps one-step demagnetization process upon laser excitation, followed by magnetization recovery which may last for hundreds of ps[4-7]. To explain this phenomenon, Beaurepaire *et al.* proposed a phenomenological three temperature (3-T) model, i.e. energy transfers among the electron, spin and lattice systems[3]. Recently, the laser-induced spin dynamics of rare earth-transition metal (RE-TM) alloys have attracted great attention due to their potential applications for spin-orbit-torque driven magnetization switching or all-optical switching[8-17]. With the antiferromagnetically-coupled sublattices, the ultrafast demagnetization for RE-TM alloys always displays an abnormal two-step characteristic and a much longer decay time, as compared to that of TM[18-22]. It was predicted that the laser induced two-step demagnetization and long decay time of RE-TM alloys arise from the 4*f* electrons of RE metals, which are buried in the deep shell and far away from the Fermi surface[23,24]. It has been reported that the typical decay time is 8 ps for Tb (orbital angular momentum $L=3$ ) and 40 ps for Gd ($L=0$), investigated by time-resolved x-ray magnetic circular dichroism[23]. What is intriguing is that, the fs laser with photon energies of both 1.55 eV and 3.1 eV could induce a continuous two-step demagnetization for Tb-TM alloys, but a discontinuous two-step demagnetization and a slightly faster recovery process for Gd-TM alloys[18,22]. As we know, the ultrafast demagnetization of Gd couldn't be excited by the fs laser because the 4*f* spins of Gd own a binding energy of about 8.4 eV compared to that of ~2.4 eV for Tb[25,26]. Moreover, it is not available for the angular momentum of Gd-TM alloys ($L=0$ for Gd) transferring directly from the 4*f* spins to lattice via spin orbit coupling. It comes to the important issues that how efficient the laser induced angular momentum transferring is and how we can manipulate the fast demagnetization for Gd-TM alloys. In particular for GdCoFe films, the strong 3*d*-5*d*6*s*-4*f* exchange interactions from intersublattices play important roles in the laser-induced ultrafast spin dynamics[11,24,27,28]. The intersublattice coupling strength $J_{RE-TM}$, which could be modulated by varying the Gd content, might be a key to accelerating the laser-induced ultrafast demagnetization time.

And to date, theoretical explanation of the two-step demagnetization occurring in the time regime of sub-ps to tens of ps is still in debate for RE-TM alloys [18,19,29]. Koopmans *et al.* employed the spin-flip scattering theory of Elloitt-Yafe (EY) type to explain the underlying mechanism, which is mainly decided by $T_c/\mu_{at}$, where $T_c$ and $\mu_{at}$ are the Curie temperature and atomic magnetic moment, respectively[19]. A small $T_c/\mu_{at}$ of the RE content in RE-TM alloys would readily predict a two-step demagnetization phenomenon, which could be understood based on the phenomenological 4-temperature (4-T) model with both $3d$ and $4f$ spins involved in the heat transferring process[29]. Apart from the unique laser-induced demagnetization process, the subsequent magnetization precession and damping behaviors can also be influenced by the RE dopants in RE-TM alloys. It has been shown that the precession frequency, effective gyromagnetic ratio and magnetic damping constant in Gd-TM alloys sharply increase and diverge around the angular momentum compensation point, which is RE composition dependent[30,31]. Nevertheless, the detailed dependences of precession and damping properties on $J_{RE-TM}$ are still ambiguous, the underlying mechanisms are deserved to explore for the future memory applications of fast switching.

In order to gain a comprehensive understanding of the laser-induced ultrafast demagnetization and magnetization precession process, in this article we show a detailed study in ferrimagnetic $Gd_x[FeCo]_{1-x}$ alloys by the time-resolved magneto-optical Kerr effect (TR-MOKE) technique, where the intersublattice exchange coupling strength $J_{RE-TM}$ are systematically modulated by varying the Gd composition of $x$. We found that the demagnetization processes of films with different $x$ could be described as only one stage (stage I) or two stages (stages I and II), and the decay times of both stages could be greatly modulated by $J_{RE-TM}$. In contrast, the precession frequency $f$, effective g-factor $g_{eff}$ and damping constant $\alpha$ that determined from the magnetization precession processes, are hardly affected by the coupling strength, except for the nearly compensated $Gd_x[FeCo]_{1-x}$ samples. These results are not only helpful to gain a better understanding of the exchange coupling interaction in multi-sublattice ferrimagnetic alloys, but also provides an alternative way to achieving effective manipulation of magnetization dynamics.

**Results and discussion:**

**The static magnetic properties of GdCoFe thin films.** Fig. 1(a) shows the typical in-plane magnetic hysteresis loops of $Gd_x(CoFe)_{1-x}$ films with various $x$. The magnetic hysteresis loops indicate that all the samples display in-plane magnetic anisotropy. With the increase of Gd composition, the saturation magnetization $M_s$ and magnetic coercivity $H_c$ display opposite nonmonotonic variation trends, as shown in Fig. 1(b). The $M_s$ value first decreases with increasing $x$, reaches almost zero around $x = 0.22$, and then increases. Considering that the sign of Kerr signals of $x = 0.22$ is opposite to that of $x = 0.24$, we can infer that the magnetic compensation point $x_{cp}$ is near $x \sim 0.23$ for our GdCoFe films, where the magnetizations of two sublattices (Gd and CoFe) are equal in magnitude and opposite in direction. More evidence could be seen from the expected significant increase of $H_c$ at $x \sim 0.23$, since the torque of the applied magnetic field is inversely proportional to the net saturation magnetization of $M_s$[32].

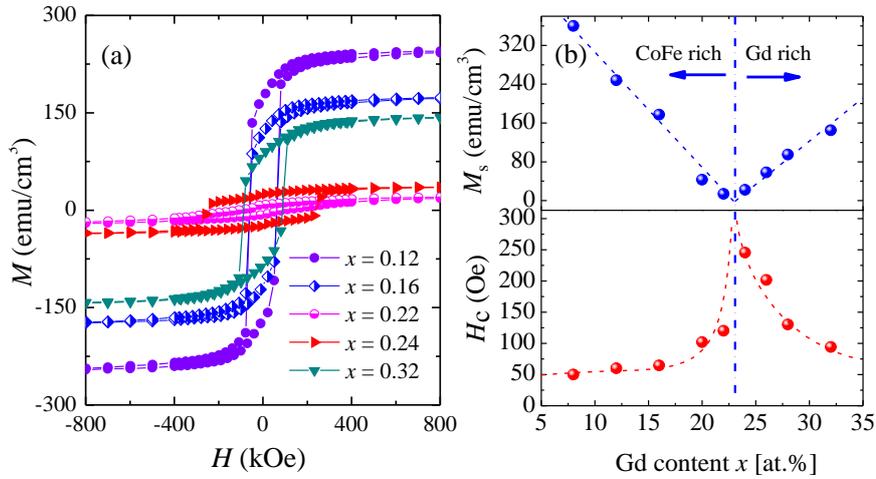

Fig. 1(a) Typical in-plane magnetic hysteresis loops for $Gd_x(CoFe)_{1-x}$ films with various $x$. (b) The saturation magnetization $M_s$ and magnetic coercivity $H_c$ as a function of Gd content $x$.

**The ultrafast demagnetization spectroscopy of GdCoFe thin films and theoretical calculations.** Figs. 2(a) and 2(b) illustrate the TR-MOKE measurement configurations of the ultrafast demagnetization and magnetization precession processes, with the magnetic field applied either in-plane or tilted with a polar angle of 19° with respect to the $Gd_x(CoFe)_{1-x}$ thin film plane, respectively. Fig. 2(c) displays several typical demagnetization curves under a saturated in-plane field of $H = 6.0$ kOe. Since the 4$f$ shell of Gd atoms is buried around 8.4 eV below $E_F$, only $s$ and $d$ electrons near the Fermi level participate in the optical excitation by the fs laser with the photon energy of 1.55 eV. The transient MOKE signals of all the GdCoFe thin films significantly vary with the pump-probe delay time $t$, and the dynamic demagnetization process could be divided into three types according to the composition of Gd. Type I of $x$=0.08~0.22 (CoFe rich) owns **a rapid one-step demagnetization within sub-ps** time range (defined as $t_1$) and a subsequent rapid recovery within several ps. For the type II samples of $x$=0.24~0.28 (Gd rich) where the sign of the transient Kerr signal reverses, two kinds of two-step demagnetization behaviors occur. Similar to the SmFe and TbFe alloys[18,33], the MOKE signal of $x$=0.24 presents a continuously decay trend which lasts for several tens of ps. In contrast, for $x$=0.26-0.28, after **the initial fast demagnetization** ($t_1$), an intermediate recovery process ($t_R$) occurs, which is then followed by **a much slow demagnetization process** ($t_2$). As for the Gd-rich samples of type III ($x = 0.32$), the varying trend is similar to that of type I, but with an opposite Kerr signal sign and a much slower recovery process.

To deeply understand the mechanism of ultrafast demagnetization for GdCoFe films, the characteristic demagnetization times of $t_1$ and $t_2$, as well as the intermediate recovery time $t_R$ were determined by using a bi-exponential fitting to the experimental data. As shown in Fig. 2(d), $t_1$ shows a nonmonotonic variation with increasing $x$. It is as low as ~ 0.5 ps for the samples with a rather low or high Gd content, which could be ascribed to the thermalization of 3$d$ electrons. However, near the compensation composition of $x$=0.22-0.28, $t_1$ is relatively larger and reaches a maximum of 1.16 ps at $x$=0.24 (close to $x_{cp}$). The decay time constant of Gd of 14 ps for 4$f$ spins and 0.8 ps for 5$d$ spins were reported, and the 3$d$-5$d$6$s$-4$f$ intersublattice exchange coupling could help the 4$f$ spins participating in the initial fast demagnetization process[24,34]. So we consider the observed larger $t_1$ is resulting from the intersublattice exchange coupling, which probably has a maximum strength at $x = x_{cp}$. Meanwhile, $t_R$ shows a similar trend to that of $t_1$ which could be ascribed to the longer demagnetization time and larger amplitude during the initial fast demagnetization process. As for the decay time $t_2$ that related to the samples of type II, it increases linearly with the Gd composition $x$, following an inverse proportion to $T_c / \mu_{at}$, as predicted by the EY scattering theory.

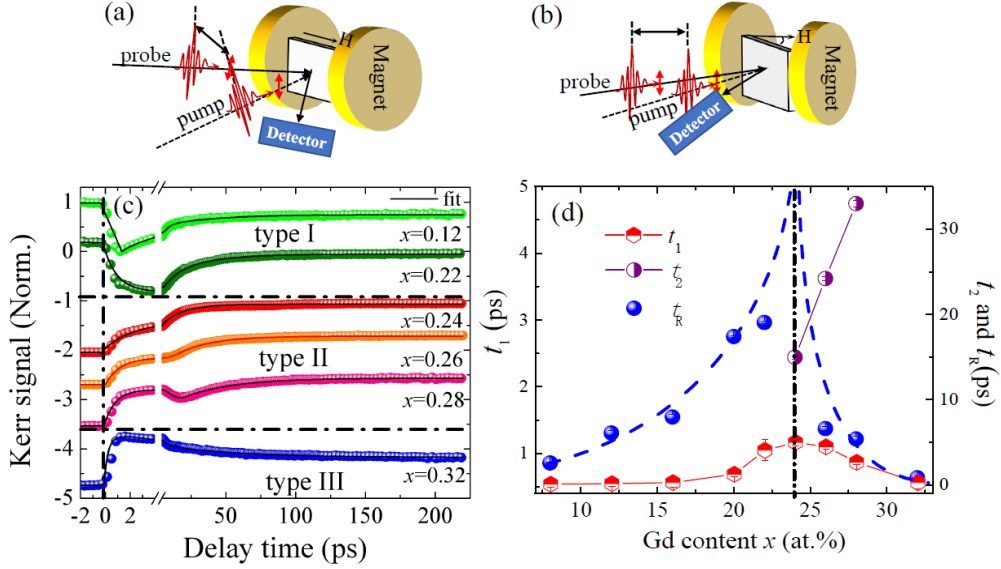

Fig. 2 TR-MOKE measurement configurations of (a) ultrafast demagnetization with $H$ applied in-plane and (b) magnetization precession with a tilted $H$. (c) The typical ultrafast demagnetization curves of $Gd_x(CoFe)_{1-x}$ films measured under a saturation in-plane magnetic field of 6.0 kOe. (d) The Gd composition dependence of characteristic times of $t_1$, $t_2$ and $t_R$.

A schematic diagram of phenomenological explanation for the demagnetization process of GdCoFe thin films could be shown in Fig. 3(a). The conduction electrons absorb the energy from the photons. At the beginning, the laser-induced angular momentum transfers from the conduction electrons to 3$d$ spins, resulting in a rapid demagnetization for all the samples. For the spin dynamics of type I, the 3$d$ spins of CoFe dominates the fast demagnetization, and the following rapid magnetization recovery could be ascribed to the electrons-phonons and 3$d$ spins - phonons coupling. For type II samples with a higher Gd content, the 5$d$6$s$ electrons serve as a heat-transfer channel for transferring the the laser-induced angular momentum to 4$f$ spins, resulting in an increase in the 4$f$ spin temperature. The heat could also returned from the 4$f$ spins to 3$d$ via the intersublattice exchange coupling, being responsible for the subsequent second demagnetization. It could be inferred that the stronger $J_{RE-TM}$ is, the longer the initial fast demagnetization time $t_1$ will be, because $t_1$ can be delayed by the excited 4$f$ spins which participate in the angular momentum transferring as well.

Based on the above analysis regarding, the temporal evolution of demagnetization are described by the 4-T model with three different 3$d$-4$f$ coupling constant $G_{ss}^{GdCoFe}$. As shown in Fig. 3(a), the 3$d$ and 4$f$ spins are regarded as two separate spin reservoirs since they have different responses to laser disturbances. By including the electronic temperature $T_e$, lattice temperature $T_l$, CoFe spin temperature $T_s^{CoFe}$, and Gd spin temperature $T_s^{Gd}$, the 4-T model is represented by four coupled differential equations, which were solved based on the Runge-Kutta method[18,29] describing heat flow between the different heat sources. The theoretically calculated results are shown in Figs. 3(b-d), which agree well with the experimental data. For all the cases, the value of $T_e$ gets to about 1020 K within ~100 fs, while the temperatures of $T_l$, $T_s^{CoFe}$ and $T_s^{Gd}$ keep almost at the room temperature. After heat transfers from electrons to other heat baths, the values of $T_l$ and $T_s^{CoFe}$ increase while $T_e$ decreases. Depending on the different values of $G_{ss}^{GdCoFe}$, there are three kinds of varying trends for the CoFe spin temperature: (i) For $G_{ss}^{GdCoFe} = 0$, $T_s^{CoFe}$ increases with the delay time $t$ to the maximum of 415 K within ~4 ps, and then decreases with further increasing $t$. Eventually, the three heat baths reach a quasi-thermal equilibrium, i.e. $T_e$,

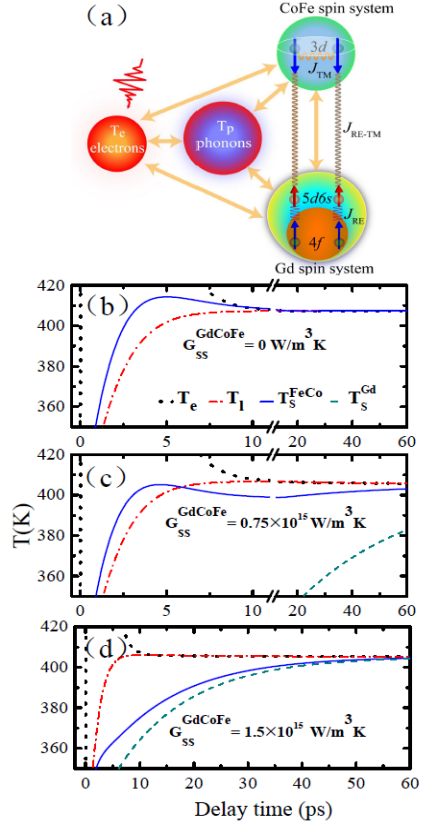

Fig. 3 (a) The Schematic diagram for the exchange coupling of the various heat baths in the 4-T model. Numerically calculated curves for the temporal evolution of the heat-reservoir temperatures ($T_e$, $T_l$, $T_s^{CoFe}$, and $T_s^{Gd}$) in the GdCoFe system using the 4-T model with $G_{ss}^{GdCoFe}=0$ (b), $G_{ss}^{GdCoFe}=0.75\times10^{15}$ W/m³K (c), $G_{ss}^{GdCoFe}=1.5\times10^{15}$ W/m³K (d).

$T_l$ and $T_s^{CoFe}$ get to a constant. In this case, $T_s^{Gd}$ does not involve in the temporal evolution of CoFe spins due to the closed heat transferring channel (without intersublattice coupling), which corresponds to the dynamic behaviors of both type I and III samples. (ii) For $G_{ss}^{GdCoFe}=0.75\times10^{15}$ W/m³K, the value of $T_s^{CoFe}$ rapidly increases within $t<4$ ps. In this time regime, $T_e$ greatly affects $T_s^{CoFe}$ via the electron-spin coupling between electrons and CoFe spins. As 4 ps $<t<$ 10 ps, accompanying with the decrease of $T_e$, $T_s^{CoFe}$ drops as well, resulting in an intermediate magnetization recovery. Finally, as $t>10$ ps, since the intersublattice coupling is strong enough, heat transferring channel between the spins of Gd and CoFe opens, and then $T_s^{CoFe}$ increases again with the increased $T_s^{Gd}$, resulting in the second demagnetization of CoFe with a longer time $t_2$. (iii) For $G_{ss}^{GdCoFe}=1.5\times10^{15}$ W/m³K, as the electron and lattice systems reach their quasi-thermal equilibrium at $t\sim10$ ps, both $T_s^{CoFe}$ and $T_s^{Gd}$ keep increasing until $t$~60 ps. This is because the large intersublattice coupling could improve the heat transferring between 4$f$ and 3$d$ spins, thus accelerating the demagnetization of Gd spins, which happens just a few ps after the CoFe spins. Such spin dynamics can be considered as a continuous two-stage demagnetization process, as compared to the discontinuous two-step demagnetization case. Apparently, since the 3$d$-4$f$ coupling constant $G_{ss}^{GdCoFe}$ is comparable with $J_{RE-TM}$, the laser-induced dynamic process of type II could be qualitatively described with an appropriate value of $J_{RE-TM}$. According to our experimental and theoretical results, it gets to a conclusion that the value of $J_{RE-TM}$ first increases with increasing the Gd composition $x$, and reaches the maximum at $x_{cp}$, then falls with further increasing $x$. The obtained dependence of $J_{RE-TM}$ on $x$ is quite different from the previous publications, which either reported a monotonic increasing trend, or a constant value[28,35].

**Field-driven ultrafast demagnetization of GdCoFe thin films.** To deeply explore the second step demagnetization of type II, the transient Kerr signals as a function of delay time were measured under various magnetic field $H$ for the samples of $x$ = 0.24 and 0.28, as shown in Figs.4 (a) and (b), respectively. The initial fast-step demagnetization time $t_1$ is field-independent and completes within ~1.5 ps, which is mainly related to the thermalization of 3$d$ spins and the influence of the partially involved 4$f$ spins on 3$d$ spins due to the strong inter-sublattice exchange coupling[24]. Nevertheless, the dynamic curves of different $H$ become separated at $t$ > 10 ps, which suggests that the second-step demagnetization process is apparently dependent on the applied field. In order to clarify this phenomenon, the characteristic time $t_2$ of $x$ = 0.24 and 0.28 and their relative amplitude ratio of the second step demagnetization $\Delta M_2$ to the total demagnetization

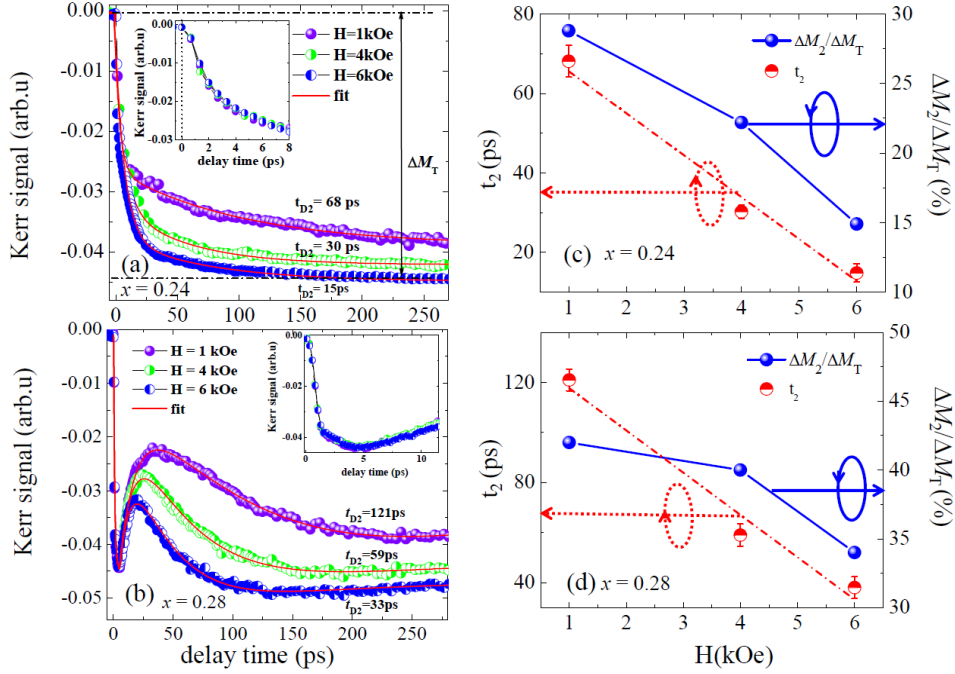

Fig. 4 The laser-induced demagnetization curves measured at various magnetic fields for the films of $x$ =0.24 (a) and 0.28 (b). The inset in (a) and (b) shows the enlarged area of delay time 12 ps. (c) and (d) demonstrate the external field dependence of $t_2$ and $\Delta M_2 / \Delta M_T$ for $x$ =0.24 and 0.28, respectively. Here $\Delta M_2$ and $\Delta M_T$ are the amplitudes of the second step demagnetization and the total demagnetization, respectively.

$\Delta M_T$, are shown in Figures 4(c) and 4(d) as a function of $H$, respectively. For both samples, $t_2$ decreases linearly with increasing $H$, which is consistent with previous reports[37]. Accompanied with the decrease of $t_2$, the ratio of $\Delta M_2 / \Delta M_T$ also shows a downward trend, suggesting that the stronger demagnetization can result in the longer $t_2$. It suggests that magnetic field could drive the second demagnetization process, ascribed to the acceleration of the growth and recovery of the reversed domains by external magnetic field[36-38]. The larger the magnetic field is, the faster the domains move. As a result, the equilibrium of demagnetization will be achieved rapidly, resulting in a small $t_2$. It could also be caused by the magnetic cooling effect, since the spin-lattice interaction can be effectively manipulated by changing the magnitude of the applied magnetic field, leading to the forced magnetization alignment in ultrafast time regime[39].

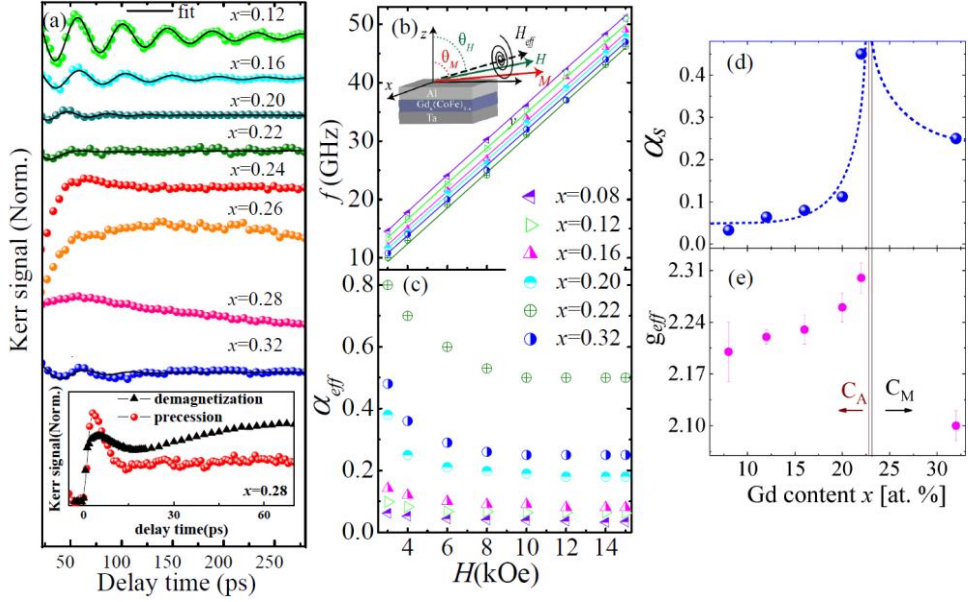

Fig. 5(a) The transient precession Kerr signals of GdCoFe films with various Gd contents and the corresponding fitting lines for $H = 6.0$ kOe. The inset in (a) shows the short range Kerr signal curves of $x = 0.28$ measured under the in-plane field (black triangles) and tilted field with a polar angle of 71 °(red circles). (b) The precession frequency $f$ as a function of $H$ (solid symbols) and the corresponding fitted lines (solid lines). (c) The effective damping factor $\alpha_{eff}$ as a function of $H$. The inset of (b) shows the polar coordinate system used for analyses of the TR-MOKE spectra. (d) and (e) show the Gd content dependence of $\alpha_s$ and $g_{eff}$ values determined from the correlation between $f$ and $H$.

**The magnetization precession and damping dynamics of GdCoFe thin films.** The magnetization precession dynamics were further measured by applying various external field $H$ with a tilted angle of $\theta_H = 19°$ with respect to the film plane, see Fig. 2(b). As we know, the magnetization vector **M** is along the direction of effective field, which mainly includes $H$ and the demagnetization field $H_d$. The variation of $H_d$ by laser heating will give rise to the precession of **M**. Fig. 5(a) shows the typical transient Kerr signals as a function of delay time measured at $H = 6.0$ kOe. Clearly, the precession curves show great changes with increasing $x$. The precession behavior gradually disappears at $x > 0.22$, and reappears at $x = 0.32$. Note that, as shown in the inset for $0.24 \leq x \leq 0.28$, all the curves show a similar two-step decay with an intermediate recovery. Especially, the inflection points of the transient Kerr signals at $t \sim 14$ ps for both the demagnetization and precession curves of $x = 0.28$ are almost the same. It could be inferred that the demagnetization of the type II dynamics has a significant effect on the precession process in the time regime of several tens of ps, since the demagnetization is not completed in the time regime.

Fig. 5(b) illustrates the polar coordinate system used for analyses of the TR-MOKE spectra, and the field dependent precession frequency $f$ curves for various $x$ derived from the Kerr signal fitting by a damped sine function [40,41]. According to the simple relation of $\alpha_{eff}= 1/2\pi f\tau$ [42], the corresponding effective magnetic damping factors are calculated and shown in Fig. 5(c). The value of $\alpha_{eff}$ decreases gradually with increasing $H$, and presents a constant as $H >10$ kOe for all the samples with various $x$. The extrinsic contribution of $\alpha_{eff}$ from local distributions of magnetization/ magnetic anisotropy could be eliminated by applying a large external magnetic field[43,44]. Therefore, we define the $\alpha_{eff}$ value at $H=12$ kOe as the intrinsic damping term $\alpha_s$. Figs. 5(d) and (e) show composition dependences of the Gilbert damping parameter $\alpha_s$ and the g-factor $g_{eff}$, which are expected to diverge at the angular momentum compensation point based on the ferromagnetic resonance mode with the LLG equations[30,45-47]. The effective gyromagnetic ratio $\gamma_{eff}$ and Gilbert damping parameter $\alpha_s$ could be expressed as follows [48].

$$\gamma_{eff}(T) = \frac{M_{CoFe}(T)-M_{Gd}(T)}{\frac{M_{CoFe}(T)}{\gamma_{CoFe}}-\frac{M_{Gd}(T)}{\gamma_{Gd}}} \quad (1),$$

$$A(T) = \frac{M_{RE}(T)}{\gamma_{RE}}-\frac{M_{TM}(T)}{\gamma_{TM}} \quad (2),$$

$$\alpha_s(T) = \frac{\alpha_{CoFe}\frac{M_{CoFe}(T)}{\gamma_{CoFe}}+\alpha_{Gd}\frac{M_{Gd}(T)}{\gamma_{Gd}}}{A(T)} \quad (3),$$

Where $M$, $\gamma$ and $\alpha$ are the net magnetic moment, gyromagnetic ratio and Gilbert damping constant of RE and TM sublattices, respectively. A(T) is the net angular momentum, $A_0$ is a constant which is independent of temperature. It is known that the g factor is ~2 for Gd and 2.16 for CoFe, which means that the angular momentum compensation point $x_{AP}$ is very close to the magnetization compensation point of $x_{CP}$ ~

0.23. Because $\alpha_s$ and $g_{eff} = \dfrac{\gamma_{eff} \cdot 2m_e c}{|e|}$ are both inversely proportional to $A(T)$, they increase firstly and then fall with increasing $x$, showing a divergence around $x \sim 0.23$. The largest $\alpha_s$ at $x \sim 0.23$ leads to the elimination of magnetization precession. In addition, for the samples with $x = 0.24$-$0.26$, the long two-step demagnetization process which lasts for several tens of ps may also be responsible for the abnormal magnetization precession curves.

**Discussion:**

Usually, 800 nm fs laser with the photon energy of 1.55eV could only excite the magneto-optical signal of GdCoFe from the Co(Fe) $3d$ moments, since the binding energy of $4f$ for Gd require photon energies about~ 8.4 eV. However, from the experimental results, two-step demagnetization could be observed due to the participation of $4f$ spins of Gd. The exchange interactions of Gd-CoFe ($J_{RE-TM}$) intersublattices play important roles in the laser-induced demagnetization dynamics.

For the CoFe-rich samples, the $3d$ moments of CoFe sublattices via ferromagnetic coupling dominate the temperature dependence of the one-step demagnetization within 1 ps. At $x \sim x_{cp}$, the strong $J_{RE-TM}$ builds up an energy transfer channel for the $3d$ and $4f$ spins. As a result, $4f$ spins could not only participate the ultrafast demagnetization within 1 ps, but also drive the second step demagnetization persisting in several tens of ps. It seems that $4f$ spins delays the ultrafast demagnetization of $3d$ spins. In other words, the $3d$-$5d6s$-$4f$ exchange interaction establishes a channel for heating up $4f$ spins. As to the Gd-rich samples with a high Gd content, the $4f$ spins are no longer involved in the ultrafast demagnetization. Since the $J_{RE-TM}$ sharply decreases due to the increase of the amount of Gd atoms, resulting in the heat transferring channel switching off between $3d$ and $4f$ spins, so the ultrafast demagnetization is dominated by $3d$ spins again. The variation trends of the first and second step decay time $t_1$ and $t_2$ with increasing $x$ could be well explained by the participation of $4f$ spins via intersublattice $3d$-$5d6s$-$4f$ exchange coupling, and the spin-flip scattering theory combining with the 4-T model, resepctively.

Moreover, we have also studied the magnetization precession processes of the GdCoFe thin films. The magnetization precession frequency $f$ and damping constant $\alpha$ could be significant modulated by the variation of $x$. Both the effective g-factor $g_{eff}$ and intrinsic damping constant $\alpha_s$ significantly increase and diverge at the angular momentum compensation point, which agree well with the ferromagnetic resonance model. These results are helpful for better understanding and controlling the magnetic dynamic behaviors of RE-TM systems with the antiferromagnetic coupling, which may find potential applications for magnetic data storage and related spintronic devices.

**Methods:**

**Magnetron sputtering of GdCoFe films on Si/SiO₂ substrates.** A series of Si /SiO$_2$ /Ta (5nm) /Gd$_x$ (Co$_{0.8}$Fe$_{0.2}$)$_{1-x}$ (20 nm) (GdCoFe) /Al (5nm) samples were fabricated by DC magnetron sputtering, with $x$ varies from 0.08 to 0.32. The base pressure of chamber was $5\times10^{-8}$ Torr and the Ar working pressure was 5.0 mTorr. The 5 nm thick Al capping layer was used in all samples to prevent from oxidation. The relative composition of Gd$_x$(CoFe)$_{1-x}$ were controlled by adjusting the deposition power of Gd target while fixing the deposition power of CoFe target and measured by energy dispersive X-Ray spectroscopy. The magnetic properties were checked by both magneto-optical Kerr effect (MOKE) and vibrating sample magnetometer (VSM). All the samples show an amorphous phase analyzed by XRD with Cu $K\alpha$ radiation.

**Time resolved Magneto-optical Kerr spectroscopy.** The fs laser pulses (duration 100 fs, center photon energy 1.55 eV) from a Ti:sapphire laser oscillator (repetition rate 1kHz) is used to excite the Gd$_x$(CoFe)$_{1-x}$ alloy films. The experimental setup schematic is shown in Fig.1. The laser-induced ultrafast demagnetization process was measured by TR-MOKE in the longitudinal geometry shown in Fig. 1(a) as a function of delay time $t$ between the pump and probe pulses, with an in-plane saturation field of 6.0 kOe applied for all the samples. In contrast, the magnetization precession and damping behavior were obtained in a polar geometry at various magnetic fields. In order to set the magnetization orientation away from the in-plane easy axis, the external magnetic field $H$ was applied at an angle of 71° with respect to the film normal, as shown in Fig. 1(b).

**Theoretical simulations and calculations of the ultrafast demagnetization.** The type I and III dynamic curves with only one-step demagnetization could be well fitted by the following equation[49-51]

$$\theta(t) = \theta_0 + H(t)\left[B\left(1 - e^{-t/t_1}\right) + C\left(1 - e^{-t/t_R}\right)\right], \quad (1)$$

where $\theta_0$ is the initial Kerr signal and $H(t)$ is the Heaviside step function. $t_1$, $t_R$, B and C are the lifetimes and amplitudes of the initial demagnetization and the following recovery, respectively. As for the type II dynamics, the two step demagnetization curves could be described by [52]

$$\theta(t) = \theta_0 + H(t)\left[B_1\left(1 - e^{-t/t_1}\right) + B_2\left(1 - e^{-t/t_2}\right)\right]. \quad (2)$$

Where $t_{1(2)}$ and $B_{1(2)}$ refer to the lifetimes and amplitudes of the demagnetization stage I and II, respectively.

The 4-T model is represented by four coupled differential equations that describe the heat flow between four heat sources, which includes electrons, CoFe spins, Gd spins and lattices. These equations can be written as follows:

$$C_e(T_e)\frac{dT_e}{dt} = -G_{el}(T_e - T_l) - G_{es}^{CoFe}(T_e - T_s^{CoFe}) - G_{es}^{Gd}(T_e - T_s^{Gd}) + P(t), \quad (3)$$

$$C_l(T_l)\frac{dT_l}{dt} = -G_{el}(T_l - T_e) - G_{ls}^{CoFe}(T_l - T_s^{CoFe}) - G_{ls}^{Gd}(T_l - T_s^{Gd}), \quad (4)$$

$$C_s^{CoFe}(T_s^{CoFe})\frac{dT_s^{CoFe}}{dt} = -G_{es}^{CoFe}(T_s^{CoFe} - T_e) - G_{ls}^{CoFe}(T_s^{CoFe} - T_l) - G_{ss}^{GdCoFe}(T_s^{CoFe} - T_s^{Gd}), \quad (5)$$

$$C_s^{Gd}(T_s^{Gd})\frac{dT_s^{Gd}}{dt} = -G_{es}^{Gd}(T_s^{Gd} - T_e) - G_{ls}^{Gd}(T_s^{Gd} - T_l) - G_{ss}^{GdCoFe}(T_s^{Gd} - T_s^{CoFe}). \quad (6)$$

Where $C_e$ is the electron specific heat and $C_l$ is the lattice specific heat. $C_s^{CoFe}$ ($C_s^{Gd}$) is the 3$d$ spins (4$f$ spins) contribution to the specific heat, respectively. $G_{el}$, $G_{es}^{CoFe}$ and $G_{es}^{Gd}$ are the electronic-lattice, electronic-spin interaction constants, respectively. $G_{ls}^{CoFe(Gd)}$ and $G_{ss}^{GdCoFe}$ are lattice-spin and 3$d$-4$f$ spins interaction constants, respectively. P(t) is the heat source term, which is only applied to the electronic subsystem. We used a laser pulse with a full width at half maximum (FWHM) of 100 femtoseconds (FWHM) and a peak power density of $3.5 \times 10^5 W/m^3$. The specific heat of the crystal lattice is taken as a constant. This approximation is available because our experiment is carried out at room temperature, and the lattice temperature is greater than the Debye temperature of the alloy. The specific heat of an electron is proportional to the temperature of the electron: $C_e(T_e) = \gamma T_e$, where $\gamma = 714 J/m^3K^2$. For CoFe sublattices, the following parameters are used: $C_s^{CoFe} = 0.13 \times 10^5 W/m^3K$, $G_{el} = 2 \times 10^{17} W/m^3K$, $G_{es}^{CoFe} = 0.13 \times 10^{16} W/m^3K$, $G_{ls}^{CoFe} = 0.48 \times 10^{16} W/m^3K$. These parameters of CoFe are close to the typical values of metals. For Gd sublattices, the following parameters are used: $C_s^{Gd} = 0.65 \times 10^5 W/m^3K$, $C_{es}^{Gd} = 0.6 \times 10^{15} W/m^3K$, $C_{ls}^{Gd} = 0.23 \times 10^{15} W/m^3K$. The 4f spin of Gd and the 3d spin of CoFe are arranged antiparallel through the 3d-5d6s-4f exchange interaction. The Runge-Kutta method is used to solve the four coupled differential equations numerically.

**The formulas for the magnetization precession.** The magnetization precession curves could be fitted by using the following formula:

$$\theta_k = a\sin(2\pi f t + \varphi)\exp(-t/\tau), \qquad (7)$$

where $a, f, \tau$ and $\varphi$ are the precession amplitude, frequency, life time, and initial phase, respectively.

The precession frequencies are well fitted by the Kittel formula [40]:

$$f = (\gamma/2\pi)\sqrt{H_1 H_2}, \qquad (8)$$

with $H_1 = H\cos(\theta_M - \theta_H) - H_{Keff}\cos 2\theta_M, \qquad (9)$

$$H_2 = H\cos(\theta_M - \theta_H) - H_{Keff}\cos^2\theta_M. \qquad (10)$$

Here $\gamma$ is the gyromagnetic ratio. $H_{Keff} = 4\pi M_s$ represents the perpendicular uniaxial magnetic anisotropy field, where $4\pi M_s$ corresponds to the demagnetization field. The magnetization orientation angle $\theta_M$ at equilibrium can be determined by the relation of $H\sin(\theta_M - \theta_H) = \dfrac{H_{Keff}}{2}\sin 2\theta_M$.